\documentclass{aa}
\usepackage{txfonts}
\usepackage{graphics,epsfig}

\newcommand{\pmn}{PMN~J0134$-$0931}

\newcommand{\kms}{km~s$^{-1}$}
\newcommand{\cm}{cm$^{-2}$}

\newcommand{\NHI}{N_{\rm HI}}

\newcommand{\lb}{\left(}

\newcommand{\rb}{\right)}
\begin{document}

\title{HI absorption in a gravitational lens at $z \sim 0.7645$}
\titlerunning{HI 21cm absorption in a gravitational lens at $z \sim 0.7645$}

\author{N. Kanekar \inst{1}\thanks{nissim@astro.rug.nl}
F. H. Briggs \inst{2}\thanks{fbriggs@mso.anu.edu.au}}
\authorrunning{Kanekar \& Briggs}
\institute{Kapteyn Institute, University of Groningen, Postbus 800, 9700 AV Groningen, 
The Netherlands \and RSAA, The Australian National University, Mount Stromlo Observatory, Cotter Road, ACT 2611, Australia}
\date{Received mmddyy/ accepted mmddyy}
\offprints{Nissim Kanekar}
\abstract{We have used the Westerbork Synthesis Radio Telescope  to
 detect HI 21cm absorption at $z \sim 0.7645$ in the gravitational lens system towards
\pmn. The 21cm profile has two broad components, with peak optical depths of
$0.047 \pm 0.007$ and $0.039 \pm 0.007$, at heliocentric redshifts $0.76470 
\pm 0.00006$ and $0.76348 \pm 0.00006$, respectively. The redshift of the 
stronger component matches that of CaII H and K absorption detected earlier. 
The absorption has a total velocity width of $\sim 500$~\kms~(between nulls) and 
an equivalent width of $7.1 \pm 0.08$~\kms. This would imply a total HI column density
of $2.6 \pm 0.3 \times 10^{21}$~\cm, for a spin temperature of 200~K
and a covering factor of unity. The high estimated HI column density is
consistent with the presence of large amounts of dust at the lens redshift; 
the intervening dust could be responsible for the extremely red colour 
of the background quasar.
}
\maketitle

\section{Introduction}

The $z \sim 2.2$ radio-loud quasar \pmn~is an exceedingly interesting object. 
Its radio continuum consists of at least 6 components, with a maximum image 
separation of $\sim 700$~mas (\cite{winn02} (W02); \cite{winn03} (W03)). Five of these
images (A -- E) have the same spectral index and a curved arc of emission is seen between 
A and B. These properties caused W02 to argue that the object is a 
gravitationally lensed system. Similarly, Gregg et al.~(2002) used the multiplicity 
of components in a near-IR image as well as the high inferred source luminosity as 
evidence for lensing. Interestingly enough, \pmn~is only the second lens of galactic 
mass to show an image multiplicity greater than four; the other such lens is the 
six-component system towards CLASS~B1359+154 (\cite{rusin01}).

CaII H and K lines have been detected towards \pmn, at $z = 0.76451 \pm 0.00016$ 
(\cite{hall02}); the lens is thus likely to be at this redshift. 
The CaII absorption is very strong, with rest frame equivalent width $\sim 5.3$
\AA, similar to that seen in the lens B0218+357 (\cite{browne93}). Further, 
the quasar is highly reddened in optical wavebands, with $B - K \ge 11$  
(\cite{gregg02}); this suggests the presence of large amounts of dust, either in 
the host galaxy of the quasar or in the lensing galaxy. While the differential 
extinction of the quasar images suggests that at least some of the reddening 
is due to dust in the absorber, it has so far not been possible to conclusively 
determine the location of the dust. Assuming the dust is at $z = 0.7645$ implies 
$E(B-V) = 1.315$, i.e. a visual extinction $A_{\rm V} = 3.85$ (\cite{hall02}). In 
conjunction with the strong CaII lines, this would suggest that the lens is a 
gas-rich, late-type galaxy.

More recently, Keeton \& Winn (2003) have come up with the first quantitative model 
of the lensing system that explains most of its observed properties. The 
background source is assumed to have two components and the lensing system 
to consist of two galaxies.
The observed 5 components A -- E are then the images of one of the background sources and 
component F is an image of the second, with the remaining two images of the 
latter source too faint for detection. The model is supported by new 
high-resolution radio and optical observations (W03), which found
that F has a different spectral index from images A -- E. These 
observations also showed that the observed large sizes of components C and E 
are not intrinsic but instead arise due to scatter broadening at the lower 
frequencies.  Finally, W03 find marginal evidence for a detection of 
the two lens galaxies in Hubble Space Telescope (HST) images of the system; Keeton \& 
Winn (2003) predict these to be spiral galaxies, each with a velocity dispersion of about 
$\sim 120$~\kms, and a projected separation of $\sim 0.4''$.

We present, in this Letter, Westerbork Synthesis Radio Telescope (WSRT) observations 
of HI 21cm absorption at $z \sim 0.7645$ towards \pmn. 
The usual association of dust with high column densities of neutral gas
confirms that the dust is likely to be concentrated in the intervening 
lens system. We describe the WSRT observations and data analysis in 
section~\ref{sec:obs}; the spectra, results and models are presented and 
discussed in section~\ref{sec:discussion}.

\section{Observations and data analysis}
\label{sec:obs}

\pmn~was observed with the UHF-High band of the WSRT on March 28 and 
31 and July 5 and 7, 2003, with the DZB correlator as the backend. 
A single 5 MHz IF band was used on all occasions, sub-divided into 
1024 channels and centred at an observing frequency of 805~MHz; this yielded 
a frequency resolution of $\sim 4.88$~kHz, i.e. a velocity resolution of 
$\sim 1.8$~\kms. The standard calibrators 3C48 and 3C147 were used to calibrate 
the absolute flux density scale at all observing epochs; their adopted flux
densities are on the scale of Baars et al.~(1977). The system bandpass was 
calibrated with observations of 3C48, 3C147, 3C454.3 or 3C138. All fourteen 
WSRT antennas were available for the observations in all observing runs. The total 
on-source time was 17.6 hours.

The data were converted from the telescope format to FITS and then analyzed 
in classic AIPS, using standard techniques. Data from the March and July 
runs were analyzed separately and self-calibrated independently. Spectra from the 
two epochs were corrected to the heliocentric frame outside AIPS and then averaged
together. The flux density of \pmn~was measured to be $590 \pm 2$~mJy on all 
observing runs.

We note that two strong, narrow RFI spikes were found to be present not far from the 
expected location of the redshifted 21cm line; these were seen in both the March and 
July runs, at the same observing frequencies (i.e. shifted heliocentric 
frequencies). 
The high initial spectral resolution of the observations limited 
the effect of these spikes on neighbouring channels; the data were also Hanning
smoothed in frequency, to further reduce spectral ringing.

\section{Discussion}
\label{sec:discussion}

\subsection{The 21cm spectrum and derived quantities}
\label{sec:21cm}

\begin{figure}
\centering
\epsfig{file=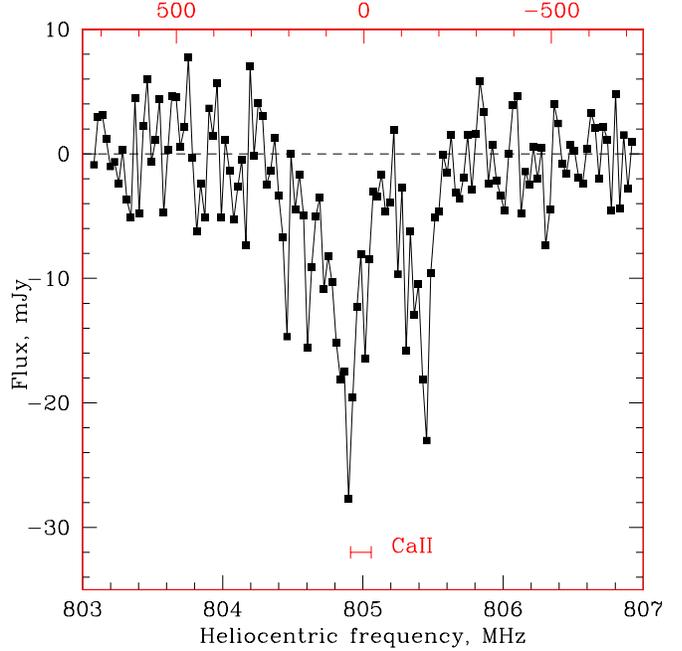,width=3.5in}
\caption{Final WSRT HI 21cm spectrum towards \pmn. The x and y axes are 
heliocentric frequency (in MHz) and flux density (in mJy) respectively; 
the top axis is heliocentric velocity (in \kms), relative to $z = 0.7645$. 
The spectrum has been boxcar smoothed by 3 channels and has a velocity 
resolution of $\sim 10.8$~\kms~and an RMS noise of $\sim 4.1$~mJy. 
}
\label{fig:lowres}
\end{figure}

The final WSRT HI 21cm spectrum towards \pmn~is shown in figure~\ref{fig:lowres}; 
this has been boxcar smoothed by three channels (and re-sampled) and has a 
velocity resolution of $\sim 10.8$~\kms~and an RMS noise of $\sim 4.1$~mJy. 
The redshift of the CaII lines (\cite{hall02}) is shown below the spectrum, 
with error bars.

Two deep features can be immediately identified in the spectrum of Fig.~\ref{fig:lowres}, 
with their peak optical depths at heliocentric frequencies of $804.898 \pm 0.03$ and 
$805.455 \pm 0.03$~MHz, i.e. heliocentric redshifts of $0.76470 \pm 0.00006$  
and $0.76348 \pm 0.00006$ respectively.
The redshift of the first of these features agrees with that of the CaII lines, within the 
errors. This is also the deeper of the two HI components, with a peak optical 
depth of $0.047 \pm 0.007$; the second feature has a peak optical depth of 
$0.039 \pm 0.007$.  The HI absorption is extremely wide, with a total velocity
spread (between nulls) of $\sim 500$~\kms. This is similar to the width of the 
CaII lines, which have $\sigma_{\rm v} = 220 \pm 40$~\kms~(\cite{hall02}).

The equivalent width of the 21cm profile is $\int \tau_{\rm 21} \mathrm{d}V = 
7.06 \pm 0.08$~\kms. This is is higher than that measured in either of 
the other gravitational lenses in which HI absorption has so far been detected,
B0218+357 ($EW = 2.94 \pm 0.02$~\kms; Kanekar et al. 2003) and PKS~1830$-$21 
($EW = 5.8 \pm 0.4$~\kms; Chengalur et al. 1999). The large equivalent width 
implies a very HI column density $\NHI = 2.6 \pm 0.3 \times 10^{21} \lb T_{\rm s} / 200 {\rm K} 
\rb \lb 1 / f \rb$~\cm~(e.g. Kanekar \& Chengalur 2003), where $f$ is the
covering factor of the absorber and $T_{\rm s}$, its spin temperature; this is well into the 
range observed in damped Lyman-$\alpha$ systems (e.g. Wolfe et al. 1986). 
We note that a spin temperature of $\sim 200$~K is typical of spiral galaxies, 
both locally (\cite{braun92}) and at low redshift (\cite{kanekar03}). 
Of course, the above estimate of $\NHI$ gives the sum of HI column densities 
towards {\it all} the different source components giving rise 
to the 21cm absorption; it is not possible to measure the column
density towards individual source components with the available data.

The extremely high estimated HI column density implies a high dust content 
in the $z \sim 0.7645$ lens and explains the observed reddening in the optical 
images (\cite{gregg02}; W03). Assuming that the molecular hydrogen column density is 
roughly equal to $\NHI$, the total hydrogen column density $N_{\rm H} = 
2 N_{\rm H_2} + \NHI \sim 7.8 \times 10^{21}$~\cm~(again assuming $T_{\rm s} = 200$~K 
and a covering factor of unity). This implies $E(B-V)\sim 1.35$, using a 
Galactic dust to gas ratio; further assuming a Galactic extinction law 
($R_{\rm V} = 3.1$; \cite{binney98}) results in a visual extinction $A_{\rm V} \sim 4.15$. 
Of course, it is likely that the 21cm absorption profile (and hence the 
estimated column density) contains contributions from more than one 
line of sight, with variable optical depth and kinematics across the 
complex lensing image; it is thus not possible to estimate the 
visual extinction towards individual source components.

\subsection{Modelling the absorption}
\label{sec:models}

\begin{figure*}
\epsfig{file=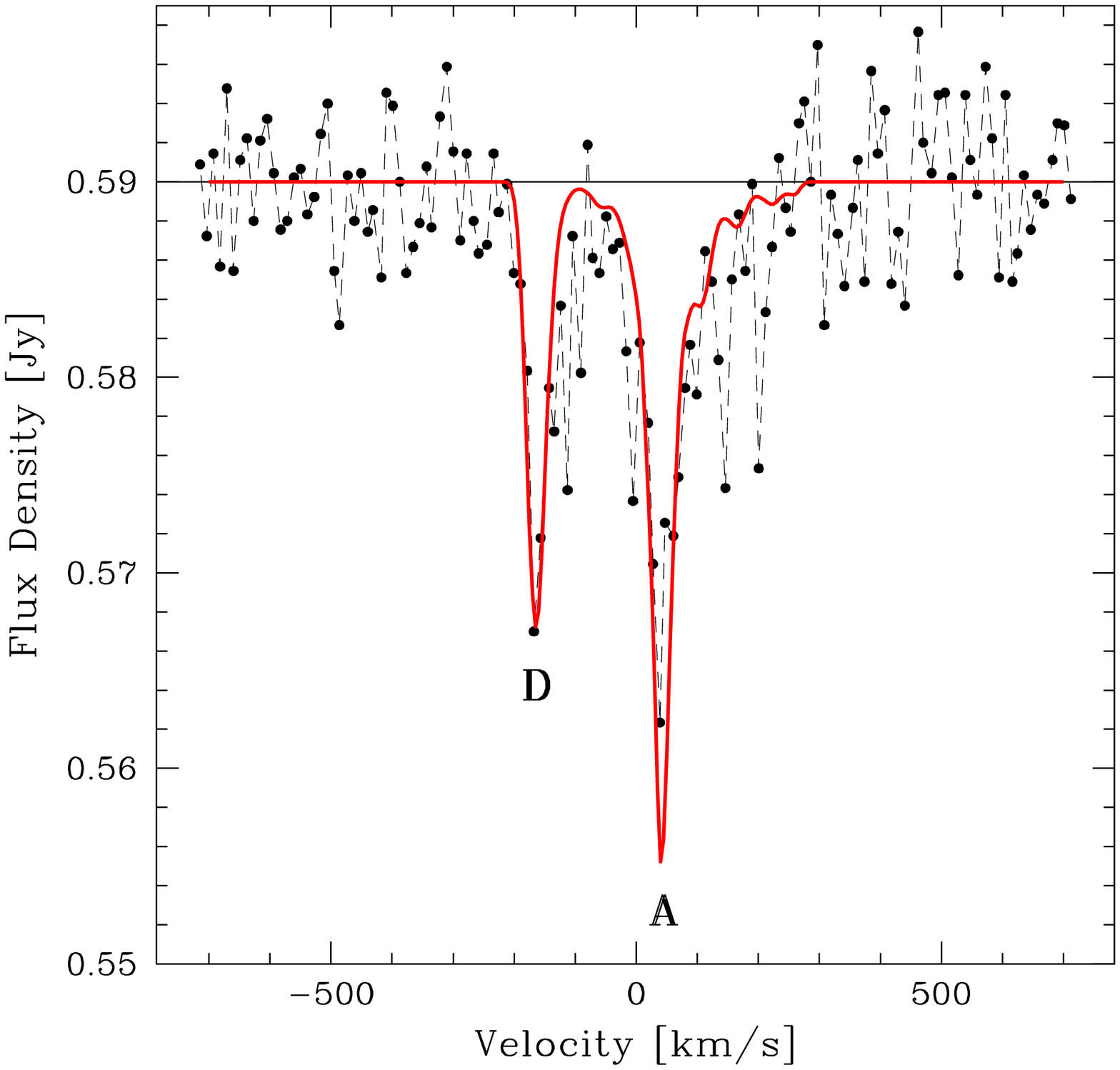,width=2.4in,height=2.4in}
\vskip -2.29in
\hskip 2.38in
\epsfig{file=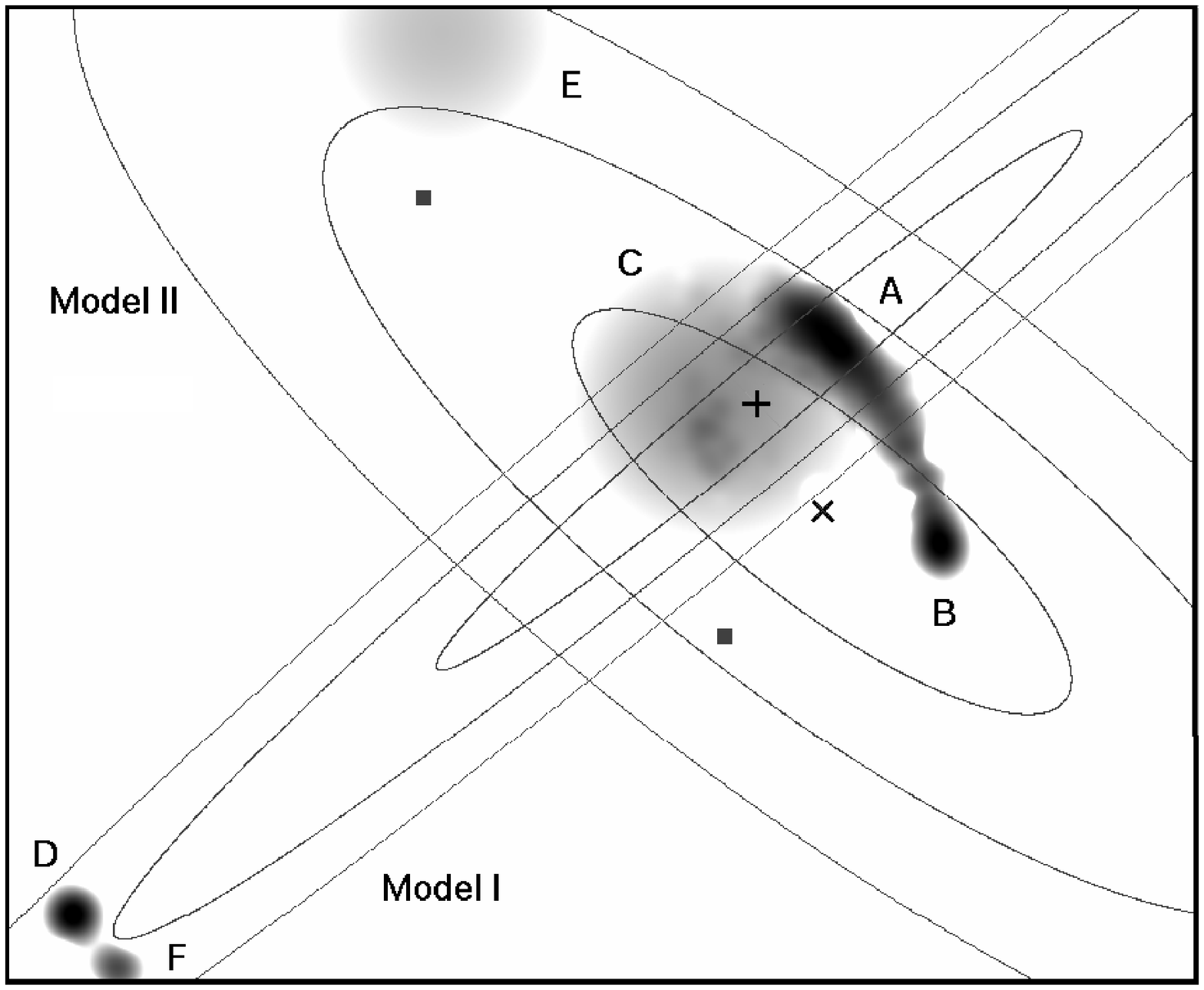,width=2.37in}
\vskip -2.1in
\hskip 4.8in 
\epsfig{file=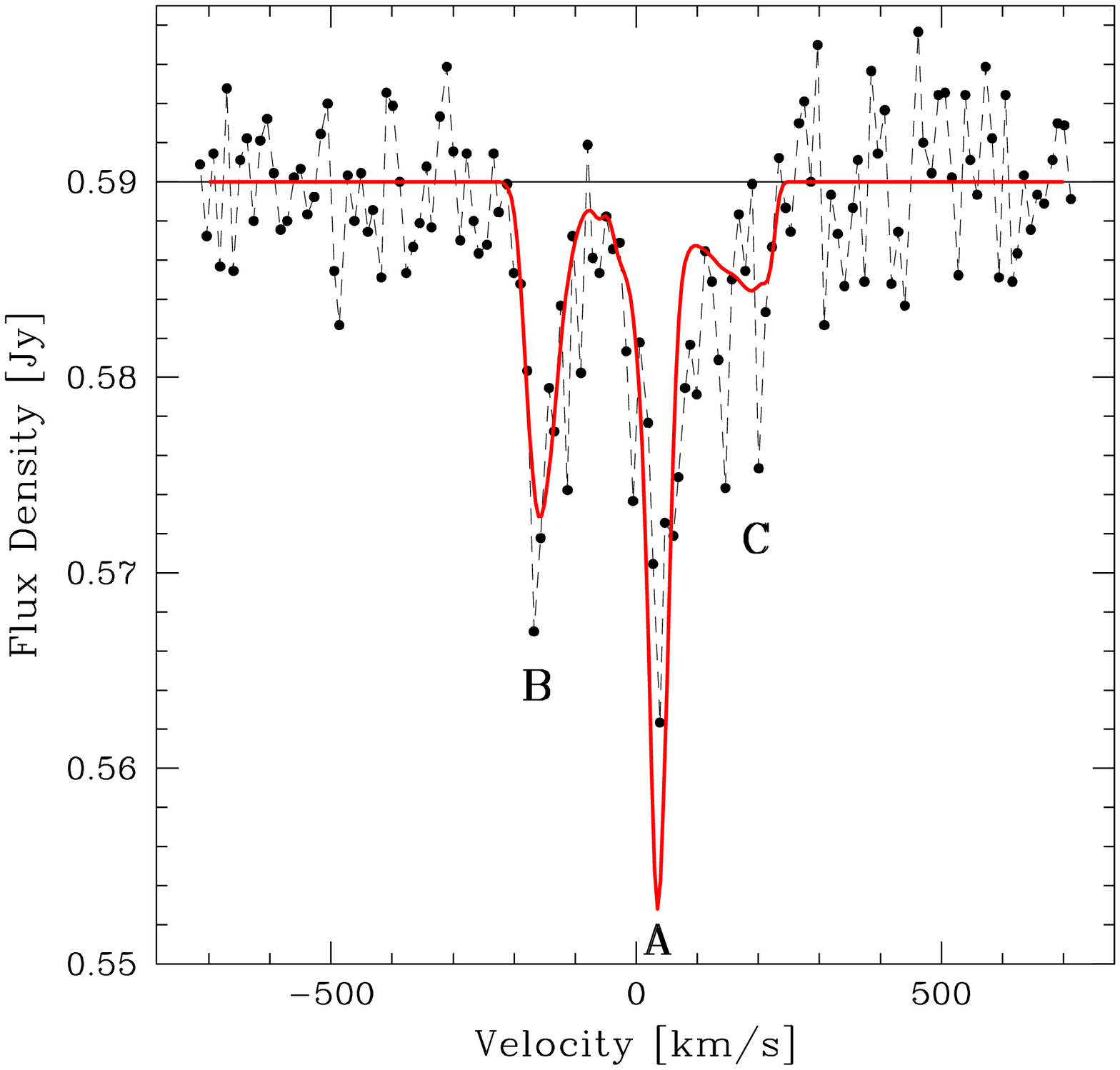,width=2.4in,height=2.4in}
\caption{Disk galaxy models for the absorption profile. Central Panel [B]:
The six principal image components (A--F) of the 1.7~GHz VLBA image of 
W03 (grey-scale), with two sets of ellipses drawn to indicate the 
orientation and inclination of the intervening model disk galaxies. 
The galaxy centre is marked by a $+$ and a $\times$ for Models~I and II, 
respectively;
the two filled squares indicate the mass centroids of the two mass
lens model of Keeton \& Winn (2003). Left Panel [A]: 21cm spectrum derived 
from Model I (heavy line) superimposed on the observed 21cm profile (dashed line). 
Right Panel [C]: 21cm spectrum derived from Model II (heavy line) superimposed 
on the observed 21cm profile (dashed line). In both cases, the zero of the velocity scale 
corresponds to 805 MHz ($z=0.76448$, heliocentric) . See text for discussion.
}
\label{fig:models}
\end{figure*}

The lens towards \pmn~creates a complex radio image of six image components (A -- F), with 
a maximum angular separation of $\sim 681$~mas (W02). Components A and B are seen 
to connect in a section of lensing arc at faint levels in the 1.7~GHz VLBA image 
(W03). Components C and E appear to be strongly scatter broadened at frequencies 
below 5~GHz and are almost undetected in VLBA images at 1.7~GHz (W02,W03), as
they are over-resolved by even the shortest baselines of the VLBA. Note, however, that 
their flux must be present in the integral flux density measured in our spectrum 
at 805~MHz; further, they provide additional lines of sight through the $z \sim 0.7645$ 
absorber. Next, while component F has a significantly steeper spectrum than the 
other five components, its total flux density is only $11 \pm 1$~mJy at 1.7~GHz; 
it is thus quite unlikely to have sufficient flux density at 805~MHz to give rise 
to either of the two main 21cm components. Further, while it is possible that the
some of the wide, weak absorption arises against this component, this would require a
fairly high optical depth ($\tau \ga 0.5$). The 21cm absorption is thus most likely to 
arise against the brighter components A, B, C and D (and perhaps the arc of radio 
emission between A and B), each of which have flux densities $\ga 100$~mJy
at 1.7~GHz. We note that component A also shows evidence for scatter broadening at
frequencies below 5~GHz while D is the most strongly reddened of the source
components; on the other hand, B is the bluest of all components and
shows no evidence for scatter broadening in the VLBA images (W03). This
suggests that there is relatively less dust and gas along the line of sight
to component B; the two main 21cm features thus appear most likely to arise
from absorption against components A and D.

We have explored these possibilities with a simple kinematic model using
a differentially rotating disk absorber. The modelling does not produce a
unique result, but rather succeeds in demonstrating that a single, massive 
galaxy could indeed be responsible for the 21cm absorption profile. If a more
complex, multi-galaxy system were to be invoked, there would be ample free 
parameters to fit the integral spectrum reported here. Further observational
progress will require radio interferometry to separate the absorption along different 
lines of sight.

The philosophy of modelling the 21cm profile shape is based on the assumption that
narrow absorption features occur in the spectrum when a compact, high-surface brightness
image of the background radio source ``selects'' a small patch from an extended, 
smooth velocity field of a rotating galaxy.  Thus, the two main narrow 21cm features 
of Fig.~\ref{fig:lowres} are expected to be associated with specific source components
of the background source, once VLBI observations are conducted in the 21~cm line.
The modelling uses the source structure obtained from the 1.7~GHz VLBA maps of W03 for 
the compact radio components, with a scaling in flux to 
both match the relative flux densities of the components in the higher frequency bands, 
and still produce the integral flux density measured at 805~MHz by our WSRT observations. 
The more extended components C and E had to be supplemented with diffuse 
gaussian components to account for the flux missing from the VLBA map.

The two models discussed here (and shown in Fig.~\ref{fig:models}[B]) are representative 
of two broad classes which can 
reproduce the observed 21cm profile. It should be emphasized, however, that neither
class is able to account for {\it all} the observations simultaneously, including the
scatter broadening and reddening of different source components. Both models are 
axially symmetric, planar
systems with rotation curves that are flat at $V_{\rm rot}$ in the outer regions, with a
linear rise from the centre to $R_{\rm c}$. The systemic velocity of the galaxy is offset 
from 805~MHz 
($z=0.76448$) by $V_{\rm off}$, and the galaxy centroid is specified by the projected 
positional offsets $\Delta X$ (west), $\Delta Y$ (north) relative to image D. The 
HI optical depth profile for each pixel is broadened by a velocity dispersion of 
$\sigma_{\rm v}=7$~km~s$^{-1}$ before computing that pixel's contribution to the integral
absorption spectrum.

\subsubsection{Model I}

Models belonging to class I are motivated by the need to inflict heavy reddening on 
image D without reddening B, coupled with accounting for the scatter broadening 
of A and C. However, this model does not account for the scatter broadening of image 
E. The model parameters are: position angle of the receding major axis $p=-50^{\circ}$ 
($+p$, measured east from north), inclination to sky plane $i=85^{\circ}$, $V_{\rm rot}=230$~km~s$^{-1}$,
$R_{\rm c}=1.4$~kpc, exponential gas scale height $h=0.7$~kpc, $V_{\rm off}= 35$~km~s$^{-1}$, 
and galaxy position at $\Delta X=490$~mas, $\Delta Y=370$~mas. In this highly 
inclined model, the HI column density measured perpendicular to the galaxy plane is only 
$3{\times}10^{19}(T_{\rm s}/100)$~cm$^{-2}$ in the region that covers components A and C, 
but it rises to $3{\times}10^{20}(T_{\rm s}/100)$~cm$^{-2}$ for B and D. The high disk 
inclination increases the line of sight column densities by an order of magnitude. 
The 21cm spectrum resulting from this model is shown in fig.~\ref{fig:models}[A].
Here, the two narrow features arise from absorption against  A and D;
the extended weak absorption wing out to $\sim +300$~\kms~is achieved by tipping
the intervening disk to a high inclination and thickening the HI layer with a 
scale height of $\sim 700$~pc (in the spirit of Briggs et al.~(1985) and 
Prochaska and Wolfe~(1997)).

\subsubsection{Model II}

The second, somewhat orthogonal, class of models (of which Model II is an example) 
has an infinitesimally thin disk, oriented to cast a shadow on images A, B, C, and E,
thereby accounting for the observed scatter broadening of A, C, and E, but implying that
all would be vulnerable to reddening in contradiction with B being the bluest and
D being the reddest of the images.  The model parameters are: $p=+52^{\circ}$,
$i=72.5^{\circ}$, $V_{\rm rot}=250$~km~s$^{-1}$, $R_{\rm c}=0.8$~kpc, (gas scale height $h=0$~kpc),
$V_{\rm off}=-15$~km~s$^{-1}$, and galaxy position at $\Delta X=535$~mas, $\Delta Y=290$~mas. 
This model has an exponential radial decline in HI column density, with scale length 
1.5~kpc, from a central peak of $1.5{\times}10^{21}(T_{\rm s}/100)$~cm$^{-2}$ measured 
perpendicular to the disk plane. The spectrum resulting from the above orientation 
is shown in fig.~\ref{fig:models}[C]. Here, the two narrow features arise against 
images A and B, while the broad shoulder at positive velocities stems from
absorption against image C and the radio arc connecting A and B.

\section{Summary}

The 21cm absorption line profile at $z \sim 0.7645$ towards \pmn~has two primary 
components and a broad weak trough, It has a total velocity width of 
$\sim 500$~\kms~and the largest HI equivalent width so far detected in a 
gravitational lens system. The extremely high estimated HI column density 
$\NHI = 2.6 \pm 0.3 \times 10^{21} \lb T_{\rm s} / 200 {\rm K} \rb \lb 1 / f 
\rb$~\cm~suggests that the reddening of the quasar is due to dust in the 
$z \sim 0.7645$ lens.

None of the galaxy models that assume a single disk-like absorber have 
succeeded in accounting for all the observations, although the models can 
roughly reproduce the observed 21cm absorption profile. Specifically, Model~I
in Sect.~\ref{sec:models} does not account for the observed scatter broadening 
of component E while Model~II fails to reproduce the fact that D is the reddest 
and B, the bluest, of all images. Of course, the large 21cm velocity width 
of $\sim 500$~\kms~could also arise if the absorption occurs in two separate 
galaxies. This is in agreement with the suggestion of Keeton \& Winn (2003) 
that the lensing system consists of two spiral systems. Further, it is 
consistent with the HST image of W03 showing marginal detections of two galaxies, 
the northern one close to images A, C and E and the southern one near 
image~D. For example, the northern ``galaxy'' in the HST image might 
cause the main 21cm feature and most of the broad, weak absorption, arising 
against images A, C, E and the radio arc, while the southern ``galaxy'' 
causes the secondary (blue-shifted) feature (and perhaps some of the broad 
trough), through absorption against image D. The present 21cm data do not 
warrant a detailed modelling of such a two-galaxy absorption system as 
the number of free parameters is far larger than the number of observational 
constraints (especially given that the observed 21cm profile can be 
reproduced with a single disk galaxy model). High resolution radio 
interferometric observations at 805~MHz will help refine the kinematic
models and shed more light on the nature of the intervening lens system.  

\begin{acknowledgements}
We thank Rene Vermeulen for much help with the planning and scheduling of these 
WSRT service observations. We are also grateful to Josh Winn for providing the 
1.7~GHz VLBA image and CC file for use in our modelling. The Westerbork Synthesis 
Radio Telescope is operated by the ASTRON (Netherlands Foundation for Research 
in Astronomy) with support from the Netherlands Foundation for Scientific Research (NWO). 
\end{acknowledgements}

\end{document}